\documentclass[11pt,twoside]{article} 
\usepackage{asp2004}
\usepackage{epsf}
\usepackage{psfig}
\usepackage{lscape} % for landscape figures/tables

\begin{document} 

\title{The White Dwarf Luminosity Function: The Shape of Things to Come}

\author{T. von Hippel,$^1$ M. Kilic,$^1$ J. Munn,$^2$ H. Harris,$^2$ K. Williams,$^3$ \\
J. Liebert,$^3$ D.E. Winget,$^1$ T.S. Metcalfe,$^4$ H. Shipman,$^5$
M.A. Wood,$^6$ T. Oswalt,$^6$ S. Kleinman,$^7$ and A. Nitta Kleinman$^7$}

\affil{$^1$The University of Texas at Austin, 1 University Station C1400,
Austin, TX 78712, USA\\
$^2$US Naval Observatory, P.O. Box 1149, Flagstaff, AZ 86002, USA\\
$^3$Steward Observatory, University of Arizona, 933 North Cherry Avenue,
Tucson, AZ 85721, USA\\
$^4$Harvard-Smithsonian Center for Astrophysics, Mail Stop 16, 60 Garden
Street, Cambridge, MA 02138, USA\\
$^5$University of Delaware, 223 Sharp Laboratory, Newark, DE 19716, USA\\
$^6$Florida Institute of Technology, Melbourne, FL  32901, USA\\
$^7$Sloan Digital Sky Survey/New Mexico State University, Apache Point
Observatory, PO Box 59, Sunspot, NM 88349, USA\\}

\begin{abstract} 
We describe a new survey for cool white dwarfs that supplements Sloan
Digital Sky Survey photometry with USNO proper motions and follow-up
spectroscopy.  To date we have discovered and spectroscopically confirmed
80 new moderate temperature and cool white dwarfs.  We have also found a
handful of high-velocity white dwarfs and we expect a sizable fraction of
these to be thick disk or possibly halo objects.  Our survey is designed
to find $\sim10^4$ new white dwarfs, although only $\sim60$ will be among
the faintest white dwarfs ($M_{\rm V} \geq 16$), where most of the
age-sensitivity resides.  We discuss an extension of our survey to $V
\approx 22$.  
\end{abstract}

\section{Introduction}

We describe a new survey for cool white dwarfs (WDs).  The goals of our
survey are to (1) place the Galactic disk, thick disk, and halo on the
same age scale to within 1 Gyr; (2) determine the age of the Galactic disk
via WD cooling ages to within 1 Gyr, well beyond the current level of
precision (e.g. Winget et al. 1987; Liebert et al. 1988; Wood 1992; Oswalt
et al. 1996; Leggett et al. 1998; Knox et al. 1999); and (3) to improve
our understanding of WD evolution.  The second goal is the most demanding,
and we anticipate working toward this for some time.  The third goal may
appear to be primarily in the domain of theory, but improved observations
are critical to progress.  For instance, the shape of the peak of the WDLF
can constrain the timing and effect on cooling of crystallization and
convective coupling of the atmosphere to the core.  Also, the shape of the
hot end of the WDLF constrains the degree of neutrino cooling in the core
of high temperature WDs (see Kim et al., these proceedings).  Of course,
theoretical and observational advances are tightly coupled, as are what we
learn from studies of cool white dwarfs and what the community learns
through WD asteroseismology (see Corsico et al. 2004; Metcalfe et al.
2004; and Fontaine \& Brassard, these proceedings).

Using the substantial photometry, astrometry, and spectroscopy from the
Sloan Digital Sky Survey (see Kleinman, these proceedings), most of the
laborious survey work has been done for us, and it is now possible to
efficiently find the cool white dwarfs that are the key to Galactic
stellar population ages.  We are supplementing SDSS photometry and
spectroscopy in such a manner as to find the cool WDs, most of which will
not otherwise be found among the SDSS data.

\section{Observational Approach}

Our first approach was to supplement the SDSS photometry with narrow-band,
DDO51 photometry, which we thought would isolate cool WDs from low
metallicity main sequence stars.  This approach turned out not to be
viable (see Kilic et al. 2004; Kilic et al., these proceedings).
Following this setback, we sought to find candidate WDs using reduced
proper motions (see Munn et al. 2005).  Proper motions were measured from
a combination of SDSS astrometry and USNO plate astrometry.  The reduced
proper motions, which combine the proper motions with SDSS photometry,
yielded hundreds of excellent candidate WDs.  We then obtained
classification-level spectroscopy for WD candidates with $g-i \geq 0$
($T_{\rm eff} \leq 8000$) at the McDonald Observatory 2.7m telescope, the
Hobby Eberly Telescope, and the MMT (see the Kilic et al.  spectroscopy
study, these proceedings).  As of the date of this conference, we had
observed 115 candidate WDs, 80 of which are WDs.

After we identify {\it bona fide} WDs from among the candidates, we will
obtain $J$-, $H$-, and $K$-band photometry, which, along with the SDSS
photometry, will yield the bolometric luminosities for these objects.
Many of the new, cool WDs we are finding have no discernible spectroscopic
features, but a subset of the most interesting WDs do have spectral
features, and we will follow these up with higher signal-to-noise
spectroscopy.  We also intend to obtain trigonometric parallaxes for as
many of the coolest WDs as possible.

\subsection{What Have We Learned So Far?}

The primary progress in our work so far has been observational, with a
technique that now produces WDs from more than two-thirds of the
candidates.  Importantly, we are now starting to understand the recovery
efficiency of WDs as a function of their location in the reduced proper
motion diagram, which leads to a new WDLF (Munn et al. 2005) based on many
more ($\geq 6000$) WDs than previously studied.  We are also finding high
velocity WDs.  These objects could be thick disk or halo objects, or they
could have been accelerated by asymmetric planetary nebulae winds
(e.g., Fellhauer et al. 2003) or binary ejection (e.g., Davies et al.
2002).  We are beginning to investigate these objects in order to
determine their parent population(s).  If these white dwarfs are thick
disk or halo objects, then we are already on our way toward good relative
ages between the major Galactic stellar populations.

\subsection{Expected Survey Yield}

We expect our survey to increase the number of WDs in the WDLF from the
current value of $\sim$400 (Liebert et al. 1988) to $\sim$10$^4$.  Because
it is always easier to find brighter representatives of any population,
most of these WDs will populate the bright and middle portion of the WDLF
(see Figure 1).  Unfortunately, the critical faint end of the WDLF, though
improved, will still be poorly populated.

%Figure 1
\begin{figure}[!t]
\plotfiddle{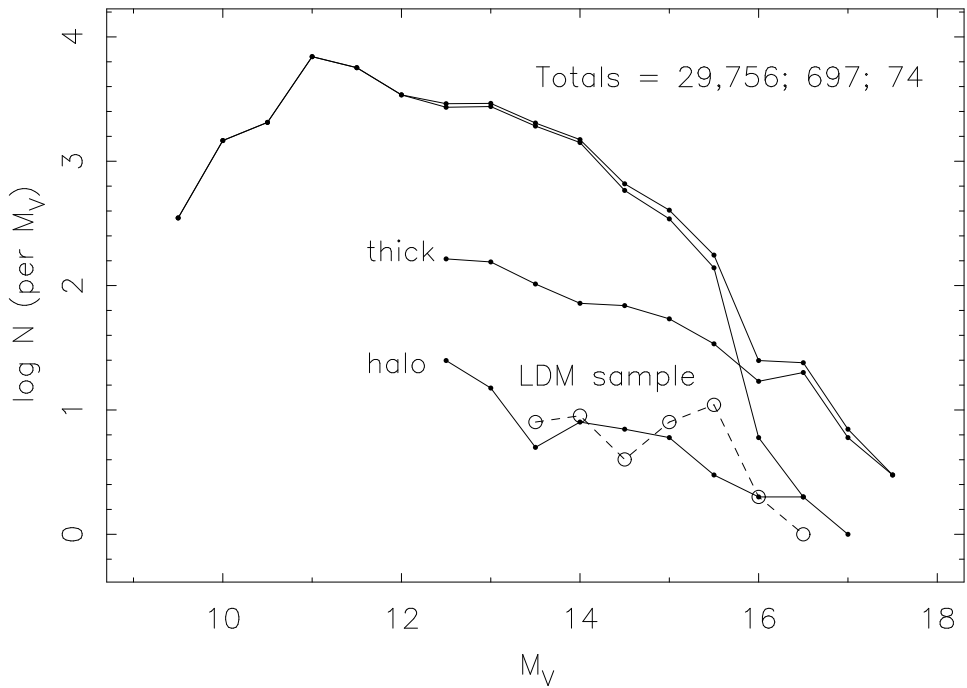}{3.6in}{270}{55}{55}{-200}{320}
\caption{Expected number of WDs found in our survey as a function of
absolute $V$ magnitude.  This estimate assumes the WDLF of Liebert et al.
(1988) for the disk, and that the thick disk and halo are single-burst, 12
Gyr old populations, with 0.2\% and 2\% of the disk normalization,
respectively.  The thick disk normalization, in particular, is
conservative and may be higher by even an order of magnitude.  This
calculation also assumes an 8000 square degree survey at a mean Galactic
latitude of 40 degrees.  This calculation incorporates only the magnitude
limit, and not the proper motion limit, so the number of the hottest WDs
will be overestimated, as many of them are too distant for a measurable
proper motion.  The total number of expected disk, thick disk, and halo
WDs, respectively, are listed in the upper right corner of the figure.}
\end{figure}

Our survey should discover hundreds of new thick disk WDs and dozens of
halo WDs.  These discoveries will allow us to make a careful comparison of
the properties of disk, thick disk, and halo WDs, and begin to work out
the relative ages of these Galactic populations.

%extra space, so makes nicer pagination
\vfill\eject

\section{The Next Step}

How do we find more of the coolest WDs where most of the age sensitivity
resides?  Our current survey is limited by the depth of the proper motion
first epoch plate material, as well as the HET and MMT spectroscopic
limit, at g $\leq$ 20.  While it is possible to obtain fainter
spectroscopy at Gemini or Keck, or with very long exposures, we need a new
astrometric epoch to push past our current survey magnitude limit, which
is set by the depth of the POSS plate material used in the USNO
astrometry.  We suggest a simple imaging survey to $r$ or $i$ = 22 over
2000--8000 square degrees.  This deep astrometric epoch would be
well-matched to the initial SDSS epoch and its filters.  In Figures 2 and
3 we show the number of WDs we expect would be found in an 8000 square
degree survey to $V$ = 21 and 22, respectively.

%Figure 2
\begin{figure}[!t]
\plotfiddle{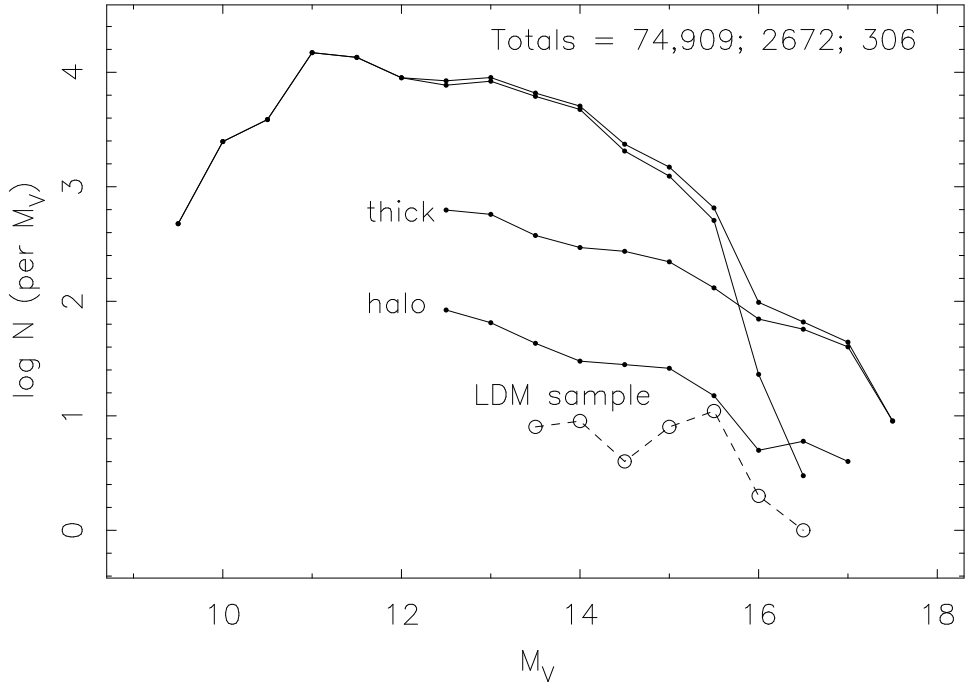}{3.6in}{270}{55}{55}{-200}{320}
\caption{Same as Figure 1, but for a survey to $V$ = 21.  This plot
over-estimates the number counts of the hottest WDs to a greater degree
than Figure 1.}
\end{figure}

%Figure 3
\begin{figure}[!t]
\plotfiddle{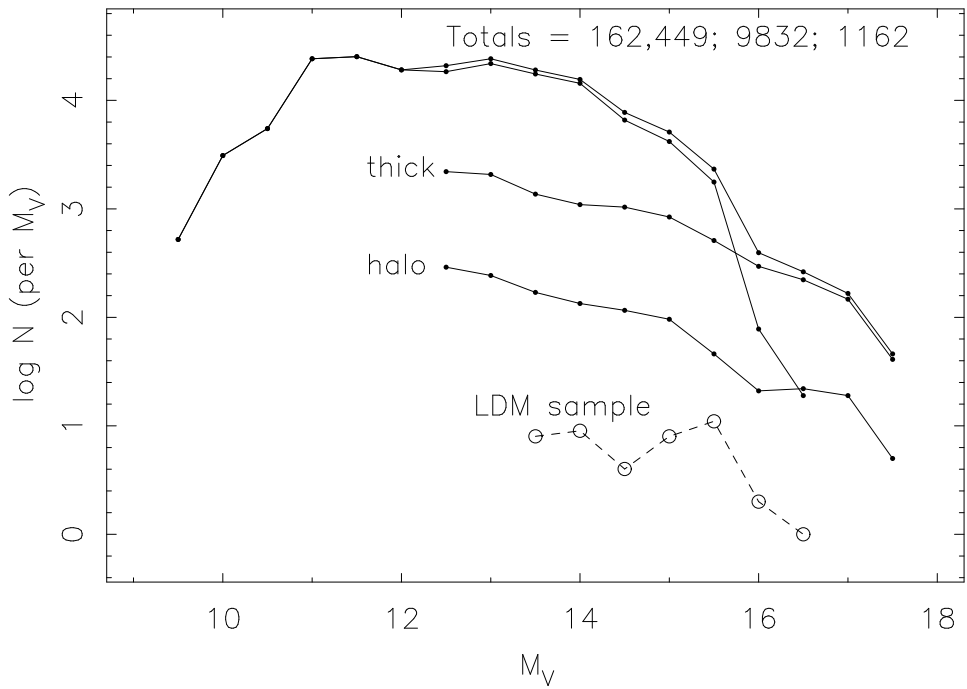}{3.6in}{270}{55}{55}{-200}{320}
\caption{Same as Figure 1, but for a survey to $V$ = 22.  This plot
over-estimates the number counts of the hottest WDs to a greater degree
than Figures 1 or 2.}
\end{figure}

\acknowledgements{TvH, MK, and DEW appreciatively acknowledge support for
this research from NASA through LTSA grant NAG5-13070 and from NSF through
grant AST-0307315.  TSM acknowledges support from the Smithsonian
Institution through a CfA Postdoctoral Fellowship, and thanks the SAO TAC
for allocating time for this project on the MMT.  TDO acknowledges the NSF
for support for this work through AST-0206115.}

%extra space, so makes nicer pagination
\vfill\eject


\begin{references}

\reference Corsico, A. H., Althaus, L. G., Montgomery, M. H.,
Garcia-Berro, E., Isern, J. 2004, \aap, in press (astro-ph/0408236)

\reference Davies, M. B., King, A., \& Ritter, H. 2002, \mnras, 333, 463

\reference Fellhauer, M., Lin, D. N. C., Bolte, M., Aarseth, S. J., \&
Williams, K. A. 2003, \apj, 595, L53

\reference Kilic, M., Winget, D. E., von Hippel, T., \& Claver, C. F.
2004, \aj, in press (astro-ph/0406424)

\reference Knox, R. A., Hawkins, M. R. S., \& Hambly, N. C. 1999, \mnras,
306, 736

\reference Leggett, S.K., Ruiz, M.T., \& Bergeron, P. 1998, \apj, 497, 294

\reference Liebert, J., Dahn, C.C., \& Monet, D.G. 1988, \apj, 332, 891

\reference Metcalfe, T. S., Montgomery, M. H., \& Kanaan, A. 2004, \apj,
605, L133

\reference Munn, J., Harris, H., Liebert, J., et al. 2005, \aj, submitted

\reference Oswalt, T.D., Smith, J.A., Wood, M.A., \& Hintzen, P. 1996,
\nat, 382, 692

\reference Winget, D.E., Hansen, C.J., Liebert, J., et al. 1987, \apj,
315, L77

\reference Wood, M.A. 1992, \apj, 386, 539

\end{references}
\end{document}